# Reply to "Comment on 'Modified quantum-speed-limit bounds for open quantum dynamics in quantum channels'"

Xin Liu,[1] Wei Wu,[1,*] and Chao Wang[2]

[1]Department of Physics, School of Science, Wuhan University of Technology（WUT）, Wuhan, China

[2]School of Engineering and Digital Arts, University of Kent, Canterbury, United Kingdom

The authors of the comment[Phys. Rev. A 97, 046101 (2018)] raise that the inconsistency in calculating some common quantum-speed-limit (QSL) bounds, which is presented in our paper [Phys. Rev. A 95, 052118 (2017)], does not exist in their paper [Phys. Rev. A 94, 052125 (2016)]. Therefore they insist that their criticism to some QSL bounds is still valid. We demonstrate all the QSL bounds mentioned in the comment are similar in essence. We also show the inconsistency which is presented in their original paper can not be used to support their criticism. Furthermore, we exhibit although the inconsistency presented in our paper is not a same one to which is presented by the authors of the comment, it exists and be unavoidable in their example. Accordingly, we believe the result in our paper is usable in numerical calculations.

In our recent article [1], the interpretation of an inconsistent estimate about some common quantum-speed-limit (QSL) bounds is given. Based on our explanation, we disagree on the conclusion presented in [2] that some common QSL bounds[3,4] do not cleave to the essence of the QSL theory. We also present a method to overcome the inconformity in numerical calculations. The authors of the comment argue that our disagreement on their criticism to some QSL bounds is problematic since we misunderstood the inconsistency they found in the estimation of the minimum time of evolution given by the QSL bounds $\tau_t^x$ (x≡op,tr,hs,quant,av). Therefore they insist that their criticism to the QSL bounds is valid.

The origin of the $\tau_t^{min}$ and the $\tau_t^{av}$ is

$$B(\rho_0,\rho_\tau) \equiv arccos(F_B(\rho_0,\rho_\tau)) \leq \int_0^\tau \sqrt{\zeta_Q(t)/4}\, dt, \qquad (1)$$

which is first presented in [5]. The left-hand of Eq. (1) denotes the geodesic's length between two given states, and the right-hand of Eq. (1) is the length of the actual evolution path between the two states. As it is explained in [2], Eq. (1) implies that the length of the geodesic connecting $\rho_0$ and $\rho_\tau$ is always shorter than the length of the actual path. The equality sign in Eq. (1) can only be achieved in the dynamical process where the length of actual evolution path between the two states is just equal to the geodesic length. The original attainable $\tau^{min}$ presented in [5] is obtained only when the equality sign in Eq. (1) is achieved (see content of [5]) and it is the actual evolution time between the two given states in this special dynamical process. It is also the shortest evolution time between the two given states among all the dynamical processes.

Then the equality sign in Eq. (1) is generalized to all dynamical processes in [2] to define

$\tau_t^{min}$, which is (Eq. (3) in [2] or the comment)

$$B(\rho_0,\rho_\tau)=\int_0^{\tau_t^{min}} \sqrt{\zeta_Q(t)/4}\,dt. \qquad (2)$$

It corresponds to the time it takes the system to travel (along the actual evolution path) the same length as the geodesic's length between the two states (this is the original description in [2]). With this generalization in all dynamical processes, however, it is obvious that whenever Eq. (2) is used in a dynamical process where the length of actual evolution path between two given states is not equal to the geodesic length to calculate the $\tau_t^{min}$, the value is unattainable in this process for $B(\rho_0,\rho_\tau)<\int_0^\tau \sqrt{\zeta_Q(t)/4}\,dt$ always meets in the process through Eq. (1). Since $\tau_t^{min}\leq t$ is established in all dynamical processes (but equality sign can only be achieved in an appropriate dynamical process), Eq. (2) can not be considered wrong. Accordingly, we think an exact description to Eq. (2) should be "an attainable QSL bound between two given states can be calculated by using Eq. (2) in a proper dynamical process, however not each value calculated with Eq. (2) in different dynamical processes is attainable." Namely, $\tau_t^{min}$ is unattainable unless Eq. (2) is used in a proper dynamical process where the length of actual evolution path is equal to the geodesic length. The attainable $\tau_t^{min}$ is not only the actual evolution time between the two states in the proper dynamical process, but also the shortest evolution time between the two states among all the dynamical processes.

Now we focus on the definition of $\tau_t^{av}$ introduced in [2]. It is written as

$$\tau_t^{av}=\frac{B(\rho_0,\rho_\tau)}{v^{av}}\leq \tau, \qquad (3)$$

where $v^{av}=(1/\tau)\int_0^\tau \sqrt{\zeta_Q(t)/4}\,dt$ and $\tau$ is the actual evolution time. Eq. (3) can also be written in form of

$$\begin{aligned}\tau_t^{av}&=\frac{B(\rho_0,\rho_\tau)}{(1/\tau)\int_0^\tau \sqrt{\zeta_Q(t)/4}\,dt}\\ &=\frac{B(\rho_0,\rho_\tau)}{\int_0^\tau \sqrt{\zeta_Q(t)/4}\,dt}\tau\\ &\leq \tau\end{aligned}. \qquad (4)$$

Obviously, in a dynamical process where the length of the actual evolution path between two given states is not equal to the geodesic's length, $\tau_t^{av}<\tau$ is always set up for $\frac{B(\rho_0,\rho_\tau)}{\int_0^\tau \sqrt{\zeta_Q(t)/4}\,dt}<1$ always meets in the process. It means the value of $\tau_t^{av}$ in this dynamical process is unattainable. On the other side, in the dynamical process where the length of the actual evolution path between the two given states is equal to the geodesic's length, by substituting Eq.

(2) into Eq. (3), we have

$$\tau_t^{av} = \frac{B(\rho_0, \rho_\tau)}{v^{av}}$$

$$= \frac{\int_0^{\tau_t^{min}} \sqrt{\varsigma_Q(t)/4}\, dt}{\frac{\int_0^{\tau_t^{min}} \sqrt{\varsigma_Q(t)/4}\, dt}{\tau_t^{min}}}, \qquad (5)$$

$$= \tau_t^{min}$$

Where $v^{av} = (1/\tau_t^{min}) \int_0^{\tau_t^{min}} \sqrt{\varsigma_Q(t)/4}\, dt$ is established in this dynamical process. Eq. (5) clearly indicates that the $\tau_t^{av}$ is attainable in this dynamical process and the value of it is also the actual evolution time $\tau_t^{min}$ between the two given states in the process. Therefore the $\tau_t^{av}$ has a similar meaning to Eq. (2) ($\tau_t^{min}$), that is "an attainable QSL bound between two given states can be calculated by using $\tau_t^{av}$ in a proper dynamical process, however not each value calculated with $\tau_t^{av}$ in different dynamical processes is attainable." In fact, all the QSL bounds mentioned in [2] have this implication because $\tau_t^x \leq t$ meets in all dynamical processes in the derivation of them, nevertheless, equality sign can only be achieved in an appropriate dynamical process. That is to say, all the QSL bounds mentioned in [2] can give the attainable bound (the shortest evolution time) between two given states among all dynamical processes, whereas the attainable value can only appear in the proper dynamical process. Obviously, it does not imply that all these QSL bounds do not cleave to the essence of the QSL theory.

In the example of [2], the actual evolution path between the initial state and the final stationary (asymptotic) state of the model is not equal to the geodesic's length between them (this fact is also presented in the last paragraph of page 1 in the comment). Therefore both values calculated by using $\tau_t^{min}$ and $\tau_t^{av}$, respectively, are unattainable values in this dynamical process. In other words, the dynamical process between these two given states in the example is not the proper process to calculate the attainable QSL bound by using $\tau_t^{min}$ or $\tau_t^{av}$. Hence the inconsistency between these unattainable values in this process can verify nothing. Here we emphasize that the finite geodesic's length (or other metrics) between two states only denotes the attainable QSL bound between the two given states should be finite (as mentioned above, the attainable QSL bound is not only the actual evolution time between the two states in the proper dynamical process, but also the shortest evolution time between the two states among all the dynamical processes), but does not mean that in each dynamical process, the evolution time between the two states is finite. Actually in the example of [2], since the final stationary (asymptotic) state can only be reached at $t \to \infty$ under its system evolution, $\tau_t^x \to \infty$ is a normal phenomenon in this dynamical process. According to above analysis, we deem that the criticism to the QSL bounds is groundless.

On the other hand, the trace distance $D(p_t, p_f)$ between the evolved state $p_t$ and the final stationary (asymptotic) state $p_f$ of the dynamical process is used to determine that the final state is reached at a finite time in the example of [2] (see Fig.2 and the second paragraph of part IV in [2]). Since $D(p_t, p_f)$ is related to the evolution path of the dynamical process ($p_t$ is the evolved state in the process) and the final stationary (asymptotic) state can only be achieved at $t \to \infty$ along the evolution path of the example, it is the exact inconformity which is presented by us in [1]. Therefore even the inconsistency presented by us is not the same one to that is presented in [2], it exists and be inevitable in the calculation. As we have mentioned in [1], the inconsistency will not happen until the limit of resolution of a calculation program is achieved (the reason leading to this is also that the finial state of the trace distance can only be reached at $t \to \infty$ in the model). Thus a modified value does not imply that the result of the calculation could have an arbitrary value depending on the machine precision, but in the dynamical process where the finial state of the calculation is only reached at $t \to \infty$ along the evolution path of the process, the numerical value in simulation is restricted by the limit of resolution of the calculation program.